\documentclass[prd,twocolumn,showpacs,preprintnumbers,amsmath,amssymb,superscriptaddress,floatfix,nofootinbib]{revtex4}
\usepackage{txfonts}
\usepackage{graphicx}
\usepackage{amsmath}
\usepackage{amsfonts}
\usepackage{amssymb}
\usepackage{color}
\usepackage{multirow}
\usepackage[colorlinks, citecolor=blue,anchorcolor=red,menucolor=red,linkcolor=red,filecolor=red,runcolor=red,urlcolor=blue,frenchlinks=red]{hyperref}

\usepackage{subfigure}

\begin{document}

\title{The  $a_0(980)$ and $f_0(980)$ in the process $D_s^+ \to K^{+} K^{-} \pi^{+}$}

\author{Jun-Ya Wang}
\affiliation{School of Physics and Microelectronics, Zhengzhou University, Zhengzhou, Henan 450001, China}

\author{Man-Yu Duan}
\affiliation{School of Physics and Microelectronics, Zhengzhou University, Zhengzhou, Henan 450001, China}

\author{Guan-Ying Wang}
\affiliation{School of Physics and Electronics, Henan University, Kaifeng 475004, China}

\author{De-Min Li}
\affiliation{School of Physics and Microelectronics, Zhengzhou University, Zhengzhou, Henan 450001, China}

\author{Li-Juan Liu}
\affiliation{School of Physics and Microelectronics, Zhengzhou University, Zhengzhou, Henan 450001, China}

\author{En Wang}
\email{wangen@zzu.edu.cn}
\affiliation{School of Physics and Microelectronics, Zhengzhou University, Zhengzhou, Henan 450001, China}

\begin{abstract}
In this work, we have investigated the process $D_s^+\to K^+ K^- \pi^+$, taking into account the contributions from the $S$-wave pseudoscalar-pseudoscalar interaction within the chiral unitary approach, and also the intermediate $\phi$ resonance.  By analyzing the BESIII and {\it BABAR} measurements, we conclude that the $f_0(980)$ state, dynamically generated from the $S$-wave pseudoscalar-pseudoscalar interaction, gives the dominant contribution close to the $K^+K^-$ threshold in the $K^+K^-$ invariant mass distribution of the decay $D_s^+\to K^+ K^- \pi^+$ in $S$-wave. On the other hand, our results imply that the lineshape adopted by BESIII and {\it BABAR} for the resonances $a_0(980)$ and $f_0(980)$ is not advisable in the fit to the data close to the $K^+K^-$ threshold.
\end{abstract}



\maketitle

\section{INTRODUCTION}
\label{sec:INTRODUCTION}
Very recently, the BESIII Collaboration has performed the amplitude analysis of the process $D_s^+ \to K^+ K^- \pi^{+}$, and reported the branching fraction $\mathcal{B}(D_s^+ \to K^+ K^- \pi^{+})=(5.47\pm 0.08\pm 0.13)\%$~\cite{Ablikim:2020xlq}, which is more precise than the measurements of the E687~\cite{Frabetti:1995sg}, CLEO~\cite{Mitchell:2009aa}, and {\it BABAR}~\cite{delAmoSanchez:2010yp} Collaborations. Due to the strong overlap of $a_0(980)\to K^+K^-$ and $f_0(980)\to K^+ K^-$, and their common quantum numbers of $J^{PC}=0^{++}$, $a_0(980)$ and $f_0(980)$ are not distinguished, and the combined state, denoted as $S(980)$, has been involved in the fit of the $S$-wave $K^+K^-$ distribution of the low mass region~\cite{Ablikim:2020xlq}. The lineshape of $S(980)$ is empirically parameterized with the following formula,
\begin{eqnarray}
A_{S(980)}&=&\frac{1}{m_0^2-m^2-im_0\Gamma_0\rho_{KK}},\label{eq:bw}\\
\rho_{KK}&=&\sqrt{1-\frac{4m_K^2}{m_{K^+K^-}^2}},
\end{eqnarray}
with $\rho_{KK}$  the correction for the Lorentz invariant phase space factor, and the mass and the width of $S(980)$ are obtained as, 
\begin{eqnarray}
m_0&=&(0.919\pm 0.006)~{\rm GeV},\\
\Gamma_{0}&=&(0.272\pm 0.040)~{\rm GeV},
\end{eqnarray}
which are not consistent with those of $a_0(980)$ or $f_0(980)$~\cite{Zyla:2020zbs}. 
In contrast, the {\it BABAR} Collaboration assumes that the $f_0(980)$ resonance gives the dominant contribution in the $S$-wave $K^+K^-$ mass distribution close to the threshold, and presents the mass and width by performing the fit with the lineshape of Eq.~(\ref{eq:bw}) for $f_0(980)$,~\cite{delAmoSanchez:2010yp}  
\begin{eqnarray}
m_0&=&(0.922\pm 0.003)~{\rm GeV},\\
\Gamma_{0}&=&(0.24\pm 0.08)~{\rm GeV},
\end{eqnarray}
which are also disagreement with those of $f_0(980)$~\cite{Zyla:2020zbs}.
The reason should be that the lineshape of Eq.~(\ref{eq:bw}) is not appropriate to represent the results that one gets with a coupled-channel unitary approach for energies close to the threshold, as discussed in Refs.~\cite{Wang:2019evy,Wang:2020elp,Wang:2020wap,Wei:2021usz}. On the other hand, the Flatt\'e parametrisation, considering the coupled-channel couplings, is perfectly fine to describe the lineshape of the resonances close to the threshold (see the `Resonances' review of Particle Data Group~\cite{Zyla:2020zbs}).

As we known, there are still many explanations for the structures of  the light scalar mesons $a_0(980)$, $f_0(980)$, and $f_0(500)$~\cite{Pelaez:2015qba,Weinstein:1990gu,Baru:2003qq,Hooft:2008we,Klempt:2007cp,Oller:1997ng,Kaiser:1998fi,Oller:1997ti,Achasov:1987ts,Achasov:2003cn,Achasov:2020fee}, and lots of the information about the charmed hadron decays, accumulated in the last decade, has provided 
an ideal platform to investigate these light scalar mesons~\cite{Liang:2016hmr,Debastiani:2016ayp,Oset:2016lyh,Wang:2020pem,Li:2020fqp,Duan:2020vye,Toledo:2020zxj,Roca:2020lyi,Ikeno:2021kzf,Ling:2021qzl,Yu:2020vlt}. In particular, the advent of the chiral unitary approach for the meson-meson interaction
 allows one to make some valuable predictions with a minimum input~\cite{Oller:1997ti,Oller:1997ng,Kaiser:1998fi,Locher:1997gr,Nieves:1999bx}, and the extracted amplitudes fulfill exactly unitary in coupled channels and all the analytical properties, in particular Flatt\'e effect.
 For instance, in Ref.~\cite{Wang:2020pem}, we have analyzed the processes $\Lambda_c\to pK^+K^-$, $p\pi^+\pi^-$ by taking into account the light scalars dynamically generated  from the  $S$-wave pseudoscalar-pseudoscalar interaction,  and  the intermediate vector mesons. Our results show that the $K^+K^-$ and $\pi^+\pi^-$ mass distributions measured by the BESIII Collaboration~\cite{Ablikim:2016tze} can be well reproduced. Later, we have proposed to study the role of $a_0(980)$ in the process $\Lambda_c\to \pi^0\eta p$ in Ref.~\cite{Li:2020fqp}. In addition, we have studied the process $D^+\to \pi^+\pi^0\eta$, and conclude that there should be 
 clear signals of $a_0(980)$ in the $\pi^+\eta$ and $\pi^0\eta$ mass distributions, respectively, which can be tested by the more precise measurements in future~\cite{Duan:2020vye}. 

Theoretically, the process $D_s^+\to K^+K^-\pi^+$ has been studied by taking into account the pseudoscalar-pseudoscalar  interaction within the chiral unitary approach in Ref.~\cite{Dias:2016gou}, where the $f_0(980)$ rather than the $a_0(980)$ is emerged in the mechanism of the $W^+$ external emission, and the {\it BABAR} measurements of the $S$-wave amplitude squared can be fair reproduced~\cite{Dias:2016gou}. However, the contribution from the $a_0(980)$ was taken into account by the BESIII Collaboration based on the fact that the process $D_s^+\to a_0(980)\pi^+ $ has been observed through $D_s^+\to \pi^+\pi^0\eta$~\cite{Ablikim:2019pit}. 
Although $a_0(980)$ couples strongly to the $K^+K^-$, it is expected that $a_0(980)$ provides a small contribution in the process $D_s^+\to K^+K^-\pi^+$, since this process mainly proceeds via the $W^+$ external emission mechanism, which is color-favored, as discussed in Ref.~\cite{Dias:2016gou}. On the other hand, the process $D_s^+\to  \pi^+\pi^0\eta$ mainly proceeds via the $W^+$ internal emission mechanism, and the rescattering of $K\bar{K}\to \eta\pi$ which will dynamically generate the $a_0(980)$~\cite{Molina:2019udw,Ling:2021qzl}.

In addition to the external $W^+$ emission mechanism, there is also the contribution from the internal $W^+$ emission for the process $D^+_s\to K^+K^-\pi^+$. The internal $W^+$ emission mechanism is not considered in Ref.~\cite{Dias:2016gou}, although this mechanism maybe give rise to the  $a_0(980)$ in the $K^+K^-$ system. 

In this work, we will investigate the process $D_s^+\to K^+K^-\pi^+$ by including the mechanisms of both the external $W^+$ emission and internal $W^+$ emission, and considering the $a_0(980)$ and $f_0(980)$ as the dynamically generated states.

This paper is organized as follows. In Sec.~\ref{sec:FORMALIISM} we will present the mechanism for the decay of $D_s^+ \to K^+ K^- \pi^{+}$, and our results and discussions will be shown in Sec.~\ref{sec:RESULTS}, followed by a short summary in the last section.

\section{FORMALISM}
\label{sec:FORMALIISM}
In this section, we present the formalism for the process $D^+_s\to K^+K^-\pi^+$, which could proceed via the $S$-wave, or through the intermediate vector $\phi$ in $P$-wave. In Subsect.~\ref{sec:swave}, we introduce the microscopic mechanism of the process in $S$-wave, by taking into account the pseudoscalar-pseudoscalar interaction within the chiral unitary approach, which will dynamically generate the scalar mesons. In Subsect.~\ref{sec:pwave}, we show the formalism for the mechanism of the intermediate vector $\phi$. 

\subsection{The mechanism of $D^+_s\to K^+K^-\pi^+$ in $S$-wave }
\label{sec:swave}

 \begin{figure}[h]
  \centering
  \subfigure[]{\includegraphics[scale=0.5]{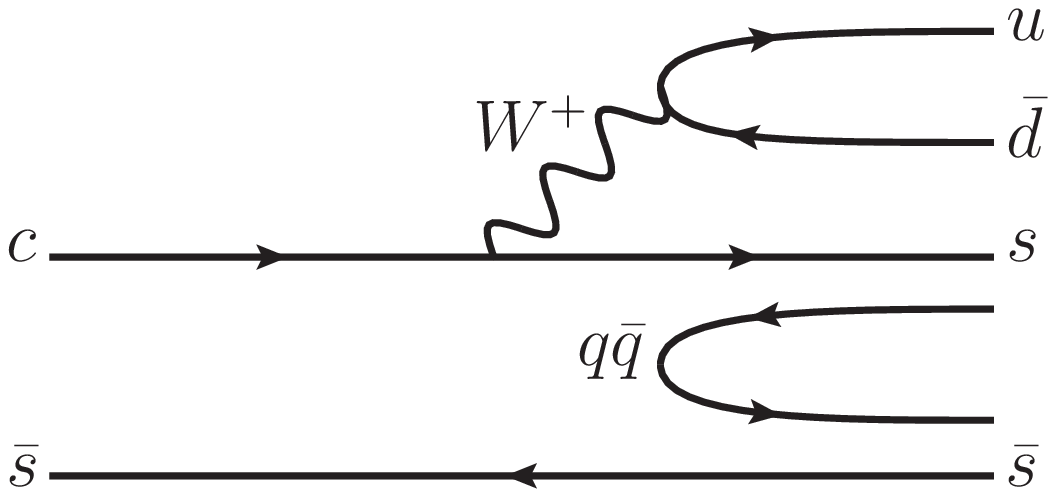}\label{fig:1a}}
  \subfigure[]{\includegraphics[scale=0.5]{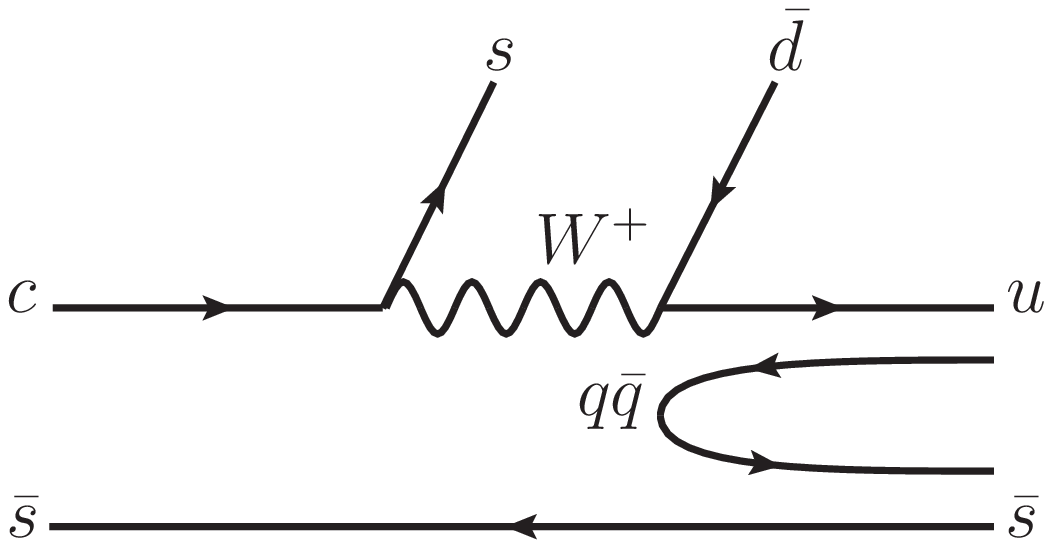}\label{fig:1b}}
  \subfigure[]{\includegraphics[scale=0.5]{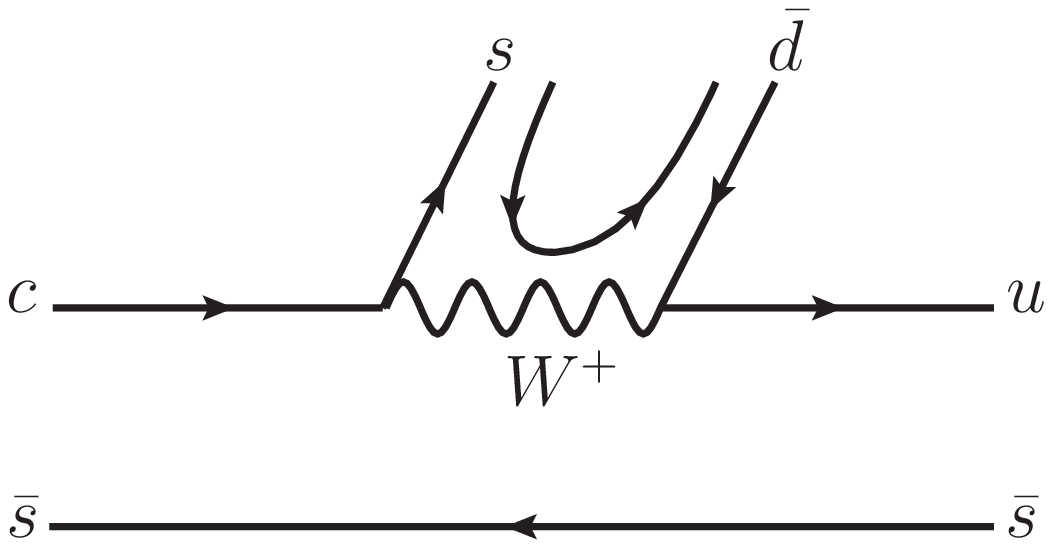}\label{fig:1c}}
  \caption{Diagrams for $D_s^+$ meson decay. (a) The $W^+$ external emission of $D_s^+\to\pi^+s\bar{s}$ and hadronization of the $s\bar{s}$ with the created $\bar{q}q$ pair, (b) the $W^+$ internal emission of $D_s^+\to\bar{K}^0u\bar{s}$ and hadronization of the $u\bar{s}$ and the created $\bar{q}q$ pair, and (c) the $W^+$ internal emission of $D_s^+\to K^+s\bar{d}$ and hadronization of the $s\bar{d}$ and the created $\bar{q}q$ pair. }
  \label{fig:quarklevel}
  \end{figure}

The Cabibbo-favored process $D_s^+ \to K^+ K^- \pi^{+}$  can happen via the weak decay of $c$ quark into a $W^+$ boson and an $s$ quark, followed by the $W^+$ boson decaying into a $u$ quark and a $\bar{d}$ quark. In order to produce the final state $K^+K^-\pi^+$,  the $s$, $\bar{d}$, and $u$ quarks from the weak decay of the $c$ quark, and the $\bar{s}$ quark of the initial $D^+_s$ meson, have to hadronize with a $q\bar{q}$ pair created from the vacuum, which can be classified as the $W^+$ external emission of Fig.~\ref{fig:quarklevel}(a), and the $W^+$ internal emission of Figs.~\ref{fig:quarklevel}(b) and \ref{fig:quarklevel}(c).

For the $W^+$ external emission  of the $D_s^+$ decay, as shown in Fig.~\ref{fig:quarklevel}(a), the $u\bar{d}$ pair forms the $\pi^+$ meson, and the remaining $s$ and $\bar{s}$ quarks, together with the created $\bar{q}q(=\bar{u}u+\bar{d}d+\bar{s}s)$ with the vacuum quantum numbers $J^{PC}=0^{++}$, hadronize into two mesons, which can be expressed as, 
\begin{eqnarray}
H^{(a)}&=&V_{cs}V_{ud}\pi^+ s\left(\bar{u} u+\bar{d} d+\bar{s}s\right)\bar{s}\nonumber\\
&=&V_{cs}V_{ud}\pi^+ \left( M_{3i}M_{i3}\right) \nonumber\\
&=&V_{cs}V_{ud}\pi^+(M^{2})_{33} ,\label{eq:quarkdiagram1}
\end{eqnarray}
where $V_{cs}$ and $V_{ud}$ are the elements of the CKM matrix. The $M$ is the $q\bar{q}$ matrix,
\begin{align}
M=\left(\begin{array}{ccc}
              u\bar{u} & u\bar{d} & u\bar{s}\\
              d\bar{u} & d\bar{d} & d\bar{s} \\
              s\bar{u} & s\bar{d} & s\bar{s}
      \end{array}
\right)\,,
\label{eq:Mmatrix}
\end{align}
 and in terms of pseudoscalar mesons the matrix $M$ can be written as~\cite{Duan:2020vye},
\begin{center}
\begin{align}
M=
\left(\begin{array}{ccc}
              \frac{\pi^0}{\sqrt{2}} +\frac{\eta}{\sqrt{3}}+\frac{\eta^\prime}{\sqrt{6}}& \pi^+ & K^+\\
              \pi^-& -\frac{\pi^0}{\sqrt{2}} +\frac{\eta}{\sqrt{3}}+\frac{\eta^\prime}{\sqrt{6}} & K^0\\
              K^-& \bar{K}^0 & - \frac{\eta}{\sqrt{3}}+\frac{2\eta^\prime}{\sqrt{6}}
      \end{array}
\right)\,.
\label{Mmatrix}
\end{align}
\end{center}
The $\eta'$ component will be ignored since the $\eta'$ has a large mass and does not play a role in the generation of $a_0(980)$ and $f_0(980)$~\cite{Oller:1997ti}. So the $(M^{2})_{33}$ of Eq.~(\ref{eq:quarkdiagram1}) can be re-expressed as
\begin{align}\label{M33}
(M^{2})_{33}&=K^+K^-+K^0\bar{K}^0+\frac{1}{3}\eta\eta.
\end{align}
Thus, we have the final states after the hadronization for the $W^+$ external emission of Fig.~\ref{fig:quarklevel}(a),
\begin{align}
\label{Ha}
H^{(a)}&=& V_{cs}V_{ud}\left(K^+K^-+K^0\bar{K}^0+\frac{1}{3}\eta\eta\right)\pi^+.
\end{align}

For the $W^+$ internal emission mechanism as depicted in Fig.~\ref{fig:quarklevel}(b), the $s$ and $\bar{d}$ quarks merge into the $\bar{K}^0$; the $u\bar{s}$ pair, together with the created $\bar{d}d$ pair, hadronizes into two mesons, 
\begin{eqnarray}\label{HaaHcc}
\label{Hb}
H^{(b)}&=&V_{cs}V_{ud}\bar{K}^0K^0\pi^+,
\end{eqnarray}
then the $\bar{K}^0K^0$ could undergo the re-scattering to $K^+K^-$. On the other hand, for another  $W^+$ internal emission mechanism of Fig.~\ref{fig:quarklevel}(c), the $u$, $\bar{s}$ quarks merge into  a $K^+$; the $s$ and $\bar{d}$ quarks, together with the created $\bar{u}u$ pair, hadronize into two mesons, 
\begin{eqnarray}
\label{Hc}
H^{(c)}&=&V_{cs}V_{ud}K^+ K^-\pi^+.
\end{eqnarray}

\begin{figure}[tbhp]
  \centering
  \subfigure[]{\includegraphics[scale=0.47]{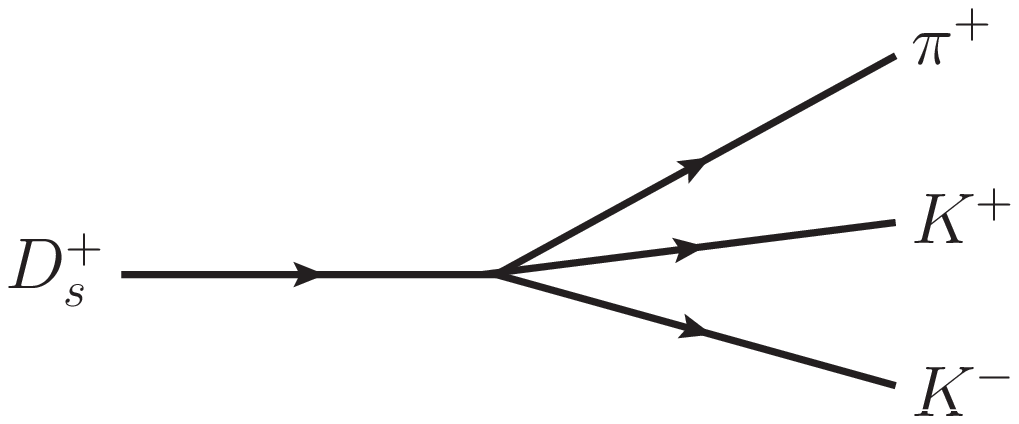}}
  \subfigure[]{\includegraphics[scale=0.47]{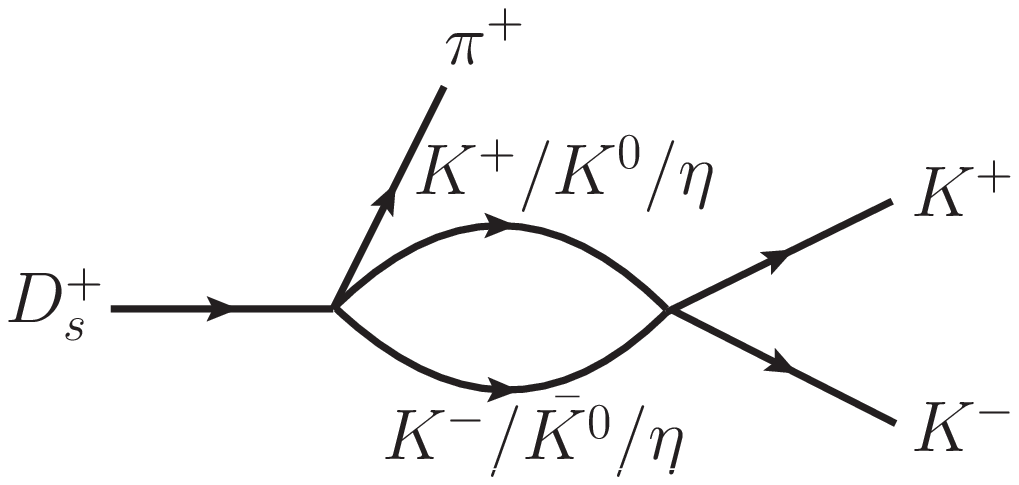}}
  \caption{\small{The $S$-wave final state interaction of the meson pairs.}}
  \label{fig:FSI}
  \end{figure}

As pointed out in Refs.~\cite{Dai:2018tgo,Duan:2020vye,Zhang:2020rqr,Dai:2018nmw}, the mechanism of the $W$ external emission is color-favored with respect to the one of the $W$ internal emission, thus, we will introduce a color factor $C$ to account for the relative weight of the $W$ external emission mechanism, and $C$ should be around 3 in the case of color number $N_c=3$. Thus, we have all the possible components after the hadronization, 
\begin{align}
H=&C H^{(a)}+H^{(b)}+H^{(c)}\nonumber\\
 =&V_{cs}V_{ud}\left\lbrace  C \left(K^+K^-+K^0\bar{K}^0+\frac{1}{3}\eta\eta\right)\pi^+ \right.
\nonumber \\ 
& \left. + \bar{K}^0K^0\pi^+  + K^+ K^-\pi^+ \right\rbrace \nonumber \\ 
= &V_{cs}V_{ud}\left\lbrace   \left(C+1\right)\left( K^+K^-\pi^++K^0\bar{K}^0\pi^+ \right) +\frac{C}{3}\eta\eta \pi^+ \right\rbrace,
\end{align}
where the coefficients of the different hadron components stand for the production weights of the corresponding components in the preliminary decay.  With the isospin doublet ($K^+$, $K^0$) and ($\bar{K}^0$, $-K^-$), we have,
\begin{eqnarray}
\left|K^+K^- \right\rangle &=& -\frac{1}{\sqrt{2}}\left( \left|K\bar{K},I=1\right\rangle + \left|K\bar{K},I=0\right\rangle \right),  \\
\left|K^0\bar{K}^0 \right\rangle &=& \frac{1}{\sqrt{2}}\left( \left|K\bar{K},I=1\right\rangle - \left|K\bar{K},I=0\right\rangle \right),
\end{eqnarray}
and,
\begin{equation}
\left|K^+K^- \right\rangle+ \left|K^0\bar{K}^0 \right\rangle =-\sqrt{2} \left|K\bar{K},I=0\right\rangle,
\end{equation}
which implies that only the isospin $I=0$ component of the $K\bar{K}$ has the contribution to the process $D_s^+\to K^+K^-\pi^+$. It should be stressed that this argument is also supported by the analysis of the {\it BABAR} Collaboration on the process $D_s^+\to K^+K^-\pi^+$~\cite{delAmoSanchez:2010yp} . 

 Indeed, for the $W^+$ external emission mechanism, the $K\bar{K}$ system is produced via the hadronization of the $s\bar{s}$ and the created $q\bar{q}$, and has the isospin $I=0$ because of the zero isospin of the $s\bar{s}$ system, as discussed in Ref.~\cite{Dias:2016gou}.
Here we have shown that the isospin $I=1$ component of $K\bar{K}$ is cancelled by summing the contributions from the $W^+$ internal emission mechanisms of Figs.~\ref{fig:quarklevel}(b) and \ref{fig:quarklevel}(c).

In addition to the $K^+K^-\pi^+$ direct production as depicted in Fig.~\ref{fig:FSI}(a), the meson pairs of $K^+K^-$, $K^0\bar{K}^0$, and $\eta\eta$ could undergo the $S$-wave final state interaction to give rise to the $K^+K^-$ final state, which will dynamically generate the scalar meson $f_0(980)$, as shown in Fig.~\ref{fig:FSI}(b).

\begin{figure}[tbhp]
  \centering
    \includegraphics[scale=0.47]{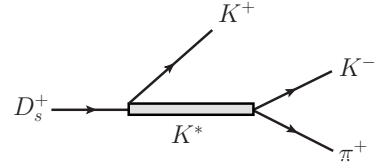}
  \caption{\small{The decay $D_s^+\to K^+K^-\pi^+$ via  the  intermediate vector $K^*$. }}
    \label{fig:Kpi}
  \end{figure}
  
\begin{figure}[htpb]
  \centering
  \includegraphics[scale=0.85]{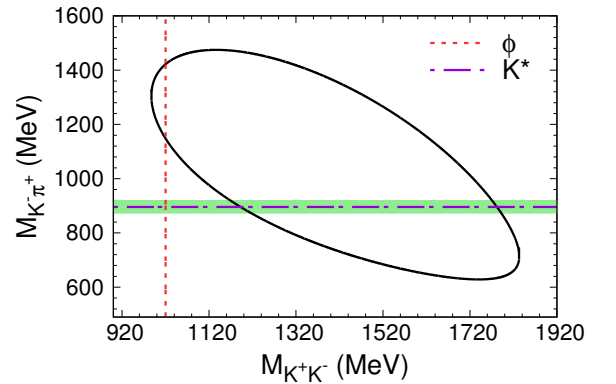}
\caption{The Dalitz plot of the decay $D_s^+ \to K^{+} K^{-} \pi^{+}$.  The colored band corresponds to the energy region ($M_{K^*}-\Gamma_{K^*}$, $M_{K^*}+\Gamma_{K^*}$).}
\label{fig:dlz12}
\end{figure}

On the other hand, the decay $D^+_s\to K^+K^-\pi^+$ can happen via  the  intermediate $K^*$, such as $K^*_0(700)$ or $K^*(892)$ ,
 as shown in Fig.~\ref{fig:Kpi}. However, the $K^*_0(700)$ and $K^*(892)$ only contribute in the regions of $1500<M_{K^+K^-}<1800$~MeV and $1200<M_{K^+K^-}<1800$~MeV, respectively, and do not affect the $K^+K^-$ invariant mass distribution close to the $K^+K^-$ threshold, which can be easily understood from the Dalitz plot shown in Fig.~\ref{fig:dlz12}.  On the other hand, the $K^*$ resonances with masses of $1100\sim 1400$~MeV ($K_1(1270)$, $K_1(1400)$, and $K^*(1410)$) may influence the structure of $K^+K^-$ invariant mass distribution around 1 GeV, since $K_1(1270)$, $K_1(1400)$, and $K^*(1410)$ have the widths larger than 100 MeV~\cite{Zyla:2020zbs}, and couple to $K\pi$ in $D$-, $D$-, and $P$-wave, respectively, we expect that their contributions are smooth and donot have the fine-structure in the $K^+K^-$ invariant mass distribution. 

Now, the amplitude of the process $D_s^+\to K^+K^-\pi^+$ in $S$-wave can be written as,
\begin{eqnarray}
{\mathcal{M}}^{S}&=&V_p\left[(C+1) + (C+1) G_{K^+K^-}t_{K^+K^-\to K^+K^-}\right.\nonumber \\
&& +(C+1)  G_{K^0\bar{K}^0} t_{K^0\bar{K}^0 \to K^+K^- }\nonumber \\
&&\left. +\frac{C}{3} G_{\eta\eta} t_{\eta\eta \to K^+K^- } \right] , \label{eq:fullamp}
\end{eqnarray}
where $V_p$ is the strength of the production vertex containing all dynamical factors.  $G_i$ is the loop function of two-meson propagator, and $t_{i\to j}$ is the transition amplitude of the $i$-channel to $j$-channel, both of which are the functions of the $K^+K^-$ invariant mass $M_{K^+K^-}$. The loop function is given by,
\begin{equation}
G_{i}= i \, \int \frac{d^4 q}{(2 \pi)^4} \,
\frac{1}{(P-q)^2 - m_1^2 + i \epsilon} \,
 \frac{1}{q^2 - m^2_2 + i \epsilon},\label{eq:loop}
\end{equation}
where $m_1$ and $m_{2}$ are the masses of the two mesons in the loop of the $i$th channel, and  $P$ and $q$ are the four-momenta of the two-meson system and the second meson, respectively. The Mandelstam invariant $s=P^2=M_{K^+K^-}^2$.
The loop function of Eq.~(\ref{eq:loop}) is logarithmically divergent, and there are two methods to solve this singular integral, either using the three-momentum cut-off method, or the dimensional regularization method. The choice of a particular regularization scheme does not, of course, affect our argumentation.  In this work, we performed the integral for $q$ in Eq.~(\ref{eq:loop}) with a cut-off $|\vec{q}_{\rm max}|= 600$ MeV~\cite{Dias:2016gou,Liang:2014tia}. The transition amplitude $t_{i\to j}$ can be obtained by solving the Bethe-Salpeter equation in coupled channels,
\begin{equation}\label{BS}
T=[1-VG]^{-1}V,
\end{equation}
where $V$ is a $5\times5$ matrix of the interaction kernel, we take five channels $\pi^+\pi^-$, $\pi^0\pi^0$, $K^+K^-$, $K^0\bar{K}^0$, and $\eta\eta$. The explicit expressions of the $5\times5$ matrix elements in $S$-wave are given by~\cite{Liang:2014tia,Ahmed:2020qkv},
\begin{eqnarray}
&& V_{11}=-\frac{1}{2f^2}s,~~~V_{12}=-\frac{1}{\sqrt{2}f^2}(s-m^2_\pi), \nonumber \\
&& V_{13}=-\frac{1}{4f^2}s,~~ V_{14}=-\frac{1}{4f^2}s,  \nonumber \\
&&V_{15}=-\frac{1}{3\sqrt{2}f^2}m^2_\pi,~~~V_{22}=-\frac{1}{2f^2}m^2_\pi,\nonumber \\
&& V_{23}=-\frac{1}{4\sqrt{2} f^2}s,~~~V_{24}=-\frac{1}{4\sqrt{2} f^2}s, \nonumber \\
&& V_{25}=-\frac{1}{6f^2}m^2_\pi,~~ V_{33}=-\frac{1}{2f^2}s,~~~V_{34}=-\frac{1}{4f^2}s,\nonumber \\
&&V_{35}=-\frac{1}{12\sqrt{2}f^2}(9s-6m^2_\eta-2m^2_\pi),\nonumber \\
&& V_{44}=-\frac{1}{2f^2}s,~~V_{45}=-\frac{1}{12\sqrt{2}f^2}(9s-6m^2_\eta-2m^2_\pi),\nonumber \\
&&V_{55}=-\frac{1}{18f^2}(16m^2_K-7m^2_\pi),
\end{eqnarray}
where $f=93$~MeV is the pion decay constant, $m_\pi$ and $m_K$ are the averaged masses of the pion and kaon, respectively~\cite{Zyla:2020zbs}.

Finally, the differential decay width for the decay $D_s^+ \to K^+ K^- \pi^{+}$  in $S$-wave is,
\begin{eqnarray}
\frac{d\Gamma}{dM_{K^+K^-}}&=&\frac{1}{(2\pi)^3} \frac{p_{\pi^+}\tilde{p}_{K^+}}{4m^2_{D_s^+}}{|{\cal {M}}^{S}|}^2.\label{eq:width}
\end{eqnarray}
where $p_{\pi}$ is the momentum of the $\pi^+$ in the $D_s^+$ rest frame, and $\tilde{p}_{K^+}$ is the momentum of the $K^+$ in the $K^+K^-$ system rest frame,
\begin{eqnarray}
p_{\pi^+}&=&\frac{\lambda^{1/2}\left(M^2_{D^+_s}, M^2_{\pi^+}, M^2_{K^+K^-}\right)}{2M_{D^+_s}},\\
\tilde{p}_{K^+}&=&\frac{\lambda^{1/2}\left(M^2_{K^+K^-}, M^2_{K^+},M^2_{K^-} \right)}{2M_{K^+K^-}}, \label{eq:ktilde}
\end{eqnarray}
with the K\"{a}llen function $\lambda(x,y,z)=x^2+y^2+z^2-2xy-2xz-2yz$.

\subsection{$D_s^+$ decays via the intermediate vector meson $\phi$}
\label{sec:pwave}
\begin{figure}[h]
  \centering
  \includegraphics[scale=0.5]{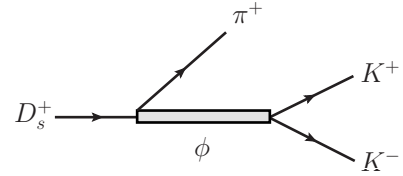}\label{fig:p1a}
\caption{The diagram for the decay of $D_s^+ \to \pi^+\phi$.}
\label{Fig:2}
\end{figure}

In this section, we will present the formalism for the decay $D_s^+ \to K^+ K^- \pi^{+}$  via the intermediate meson $\phi$. The $\phi$ could decay to $K^+K^-$ in $P$-wave. The quark level diagram is shown in Fig.~\ref{Fig:2} where the $D_s^+$  decays into the meson $\pi^+$ and the vector meson $\phi$.

Since the contributions of $S$-wave and $P$-wave add incoherently, we can rewrite the differential decay width of Eq.~(\ref{eq:width}) by including the $P$-wave contribution as,
\begin{align}
\frac{d\Gamma}{dM_{K^+K^-}}&=&\frac{1}{(2\pi)^3} \frac{p_{\pi^+}\tilde{p}_{K^+}}{4m^2_{D_s^+}}\left( |{\cal {M}}^{S}|^2 +|{\cal {M}}^{P}|^2\right),\label{eq:totalwidth}
\end{align}
and the amplitude of the $P$-wave is given by~\footnote{The details is given in Appendix~\ref{app:pwave}.},
\begin{eqnarray}
{\mathcal{M}}^{P}&=&\frac{\beta \tilde{p}_{\pi^+} \tilde{p}_{K^+}}{M^2_{K^+K^-}-M^2_{\phi}+i M_{\phi} \Gamma_{\phi}}.
\end{eqnarray}
where $\beta$ is the strength of contribution from the term of the intermediate $\phi$. $\tilde{p}_{\pi^+}$ is the momentum of the $\pi^+$ in the $K^+K^-$ rest frame, and $\tilde{p}_{K^+}$ is the momentum of the $K^+$ in the $K^+K^-$ system rest frame
\begin{eqnarray}
\tilde{p}_{\pi^+}&=&\frac{\lambda^{1/2}\left(M^2_{D^+_s}, M^2_{\pi^+}, M^2_{K^+K^-}\right)}{2M_{K^+K^-}},\\
\tilde{p}_{K^+}&=&\frac{\lambda^{1/2}\left(M^2_{K^+K^-}, M^2_{K^+},M^2_{K^-} \right)}{2M_{K^+K^-}}.
\end{eqnarray}
The masses of the mesons are taken from the Particle Data Group (PDG)~\cite{Zyla:2020zbs}.

\section{RESULTS AND DISCUSSIONS}
\label{sec:RESULTS}

\begin{figure}[htpb]
  \centering
  \includegraphics[scale=0.5]{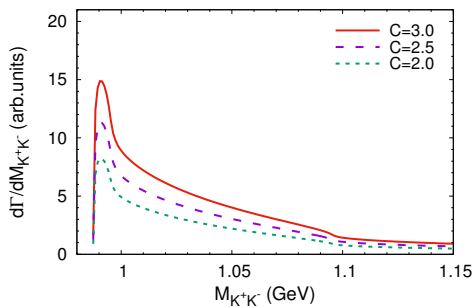}
  \caption{\small{The $K^+K^-$ invariant mass distributions for the $D_s^+ \to K^+ K^- \pi^{+}$ decay with different values of color factor $C$.}}
  \label{fig:cc}
  \end{figure}

In our theoretical model, we have three parameters: (1), the normalization factor $V_p$ of Eq.~(\ref{eq:fullamp}); (2), the color factor $C$; and (3), $\beta$ the strength of the contribution from the intermediate $\phi$.  
Since the color factor $C$ should be around 3,  in Fig.~\ref{fig:cc} we show the $K^+K^-$ mass distribution for the $D_s^+\to K^+K^- \pi^+$ in $S$-wave with the $C=3.0$, $2.5$, $2.0$ up to an arbitrary normalization. One can easily find that the value of $C$ does not affect the lineshape of $S$-wave contribution. Thus, we will fit the parameters $V_p$ and $\beta$ to the experimental data of the BESIII and {\it BABAR} Collaborations by taking $C=3$.

It should be pointed out that the distributions of $|S|^2$ and $|P|^2$ of BESIII~\cite{Ablikim:2020xlq} and {\it BABAR}~\cite{delAmoSanchez:2010yp}, regarded as the amplitude squared $|t_{K^+K^-}|^2$ in Ref.~\cite{Dias:2016gou}, turn out to be the $K^+K^-$ mass distributions of the decay $D_s^+\to K^+K^-\pi^+$ in $S$-wave and $P$-wave, respectively. We  give the explicit explanation in Appendix~\ref{app:B}.

\begin{figure}[htpb]
  \centering
  \includegraphics[scale=0.5]{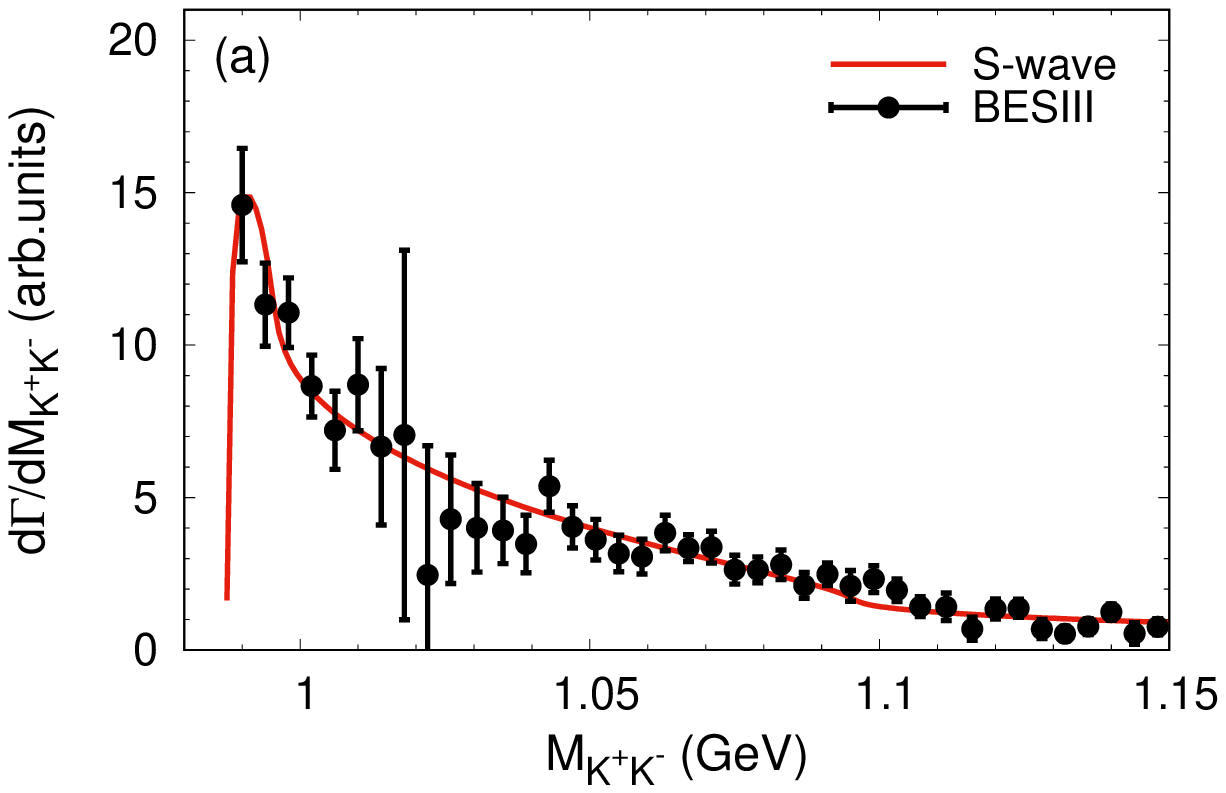}
  \includegraphics[scale=0.5]{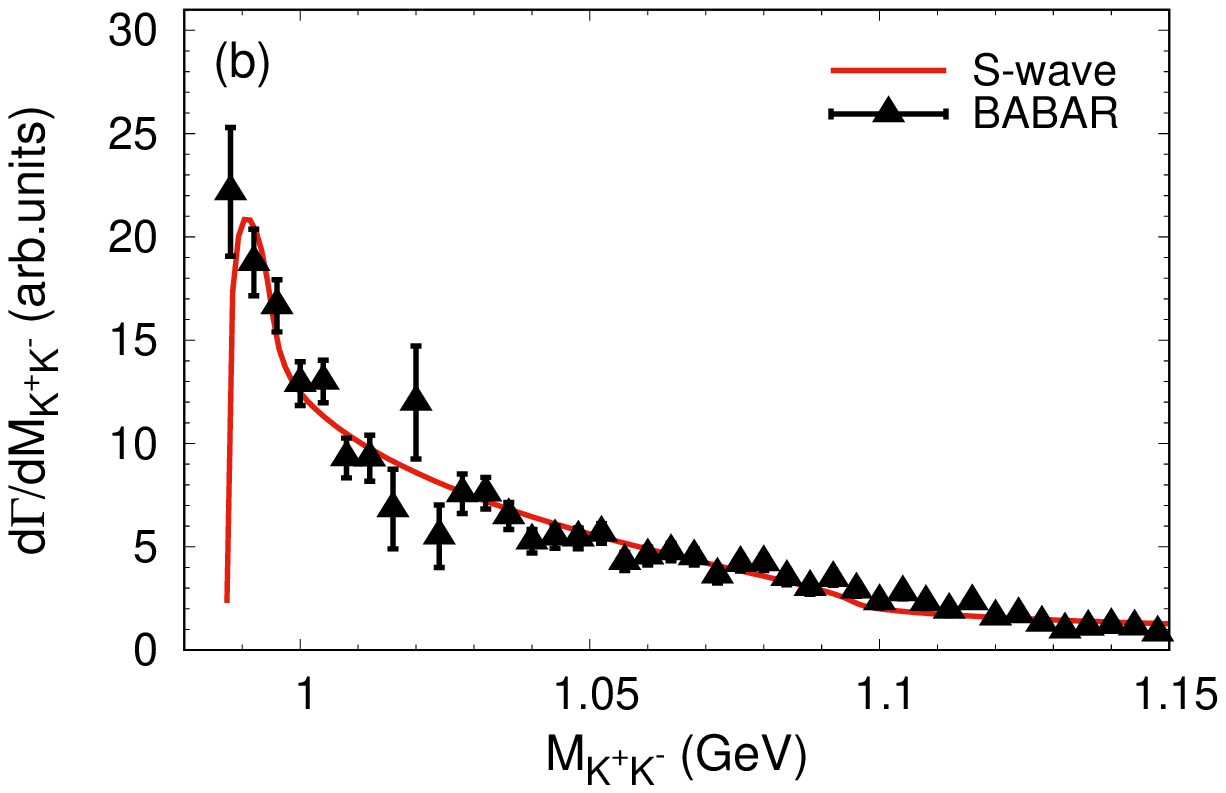}
  \caption{\small{The $K^+K^-$ mass distributions of the decay $D_s^+\to K^+K^- \pi^+$ in $S$-wave with the fitted parameters of Table~\ref{table:p}. The BESIII data are taken from Fig.~9(a) of Ref.~\cite{Ablikim:2020xlq}, and the {\it BABAR} data are taken from Fig.~5(b) of Ref.~\cite{delAmoSanchez:2010yp}.}}
  \label{fig:ss}
  \end{figure}

Firstly, we fit the $V_p$ to the BESIII and {\it BABAR} measurements of the $K^+K^-$ mass distribution of the decay $D_s^+\to K^+K^-\pi^+$ in $S$-wave~\cite{Ablikim:2020xlq,delAmoSanchez:2010yp}. The $\chi^2/d.o.f$ is $35.66/(40-1)$ and $87.29/(40-1)$, respectively for the BESIII and {\it BABAR} data. Our results calculated with the fitted parameters tabulated in Table~\ref{table:p} are shown in Fig.~\ref{fig:ss}. One can find that our results, involving the $S$-wave pseudoscalar-pseudoscalar interaction within the chiral unitary approach, are in good agreement with the BESIII and {\it BABAR} measurements. We can conclude that the enhancement near the $K^+K^-$ threshold is mainly due to the $f_0(980)$, and the combined state $S(980)$ with the lineshape of Eq.~(\ref{eq:bw}) used by the BESIII and {\it BABAR} is not suitable.

\begin{table*}
\begin{center}
\caption{ \label{table:p} The parameters fitted to the BESIII~\cite{Ablikim:2020xlq} and {\it BABAR}~\cite{delAmoSanchez:2010yp} data.}
\begin{tabular}{c  c  c  |  c c}
\hline\hline
 parameters   &$V_{p}$       &$\chi^2/dof$          & $\beta$      &$\chi^2/dof$ \\
\hline
  BESIII       & $2540.9\pm 0.4$   &  $35.66/(40-1)$         & $2696.5\pm 0.55$    & $316.24/(40-1)$  \\
   {\it  BABAR}       & $3008.3\pm 0.4$   &  $87.29/(41-1)$       & $3233.3\pm 0.60$    & $1137.62/(41-1)$  \\
\hline\hline
\end{tabular}
\end{center}
\end{table*}

  \begin{figure}[htpb]
  \centering
  \includegraphics[scale=0.5]{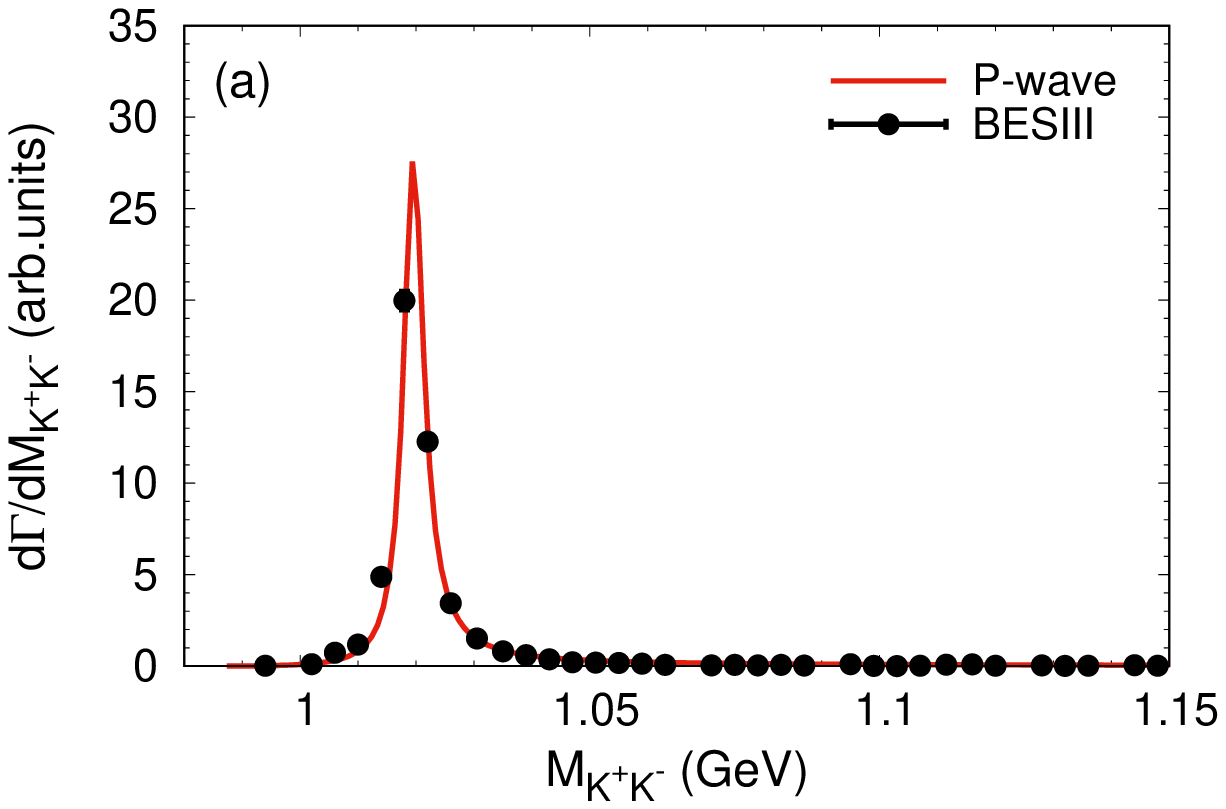}
  \includegraphics[scale=0.5]{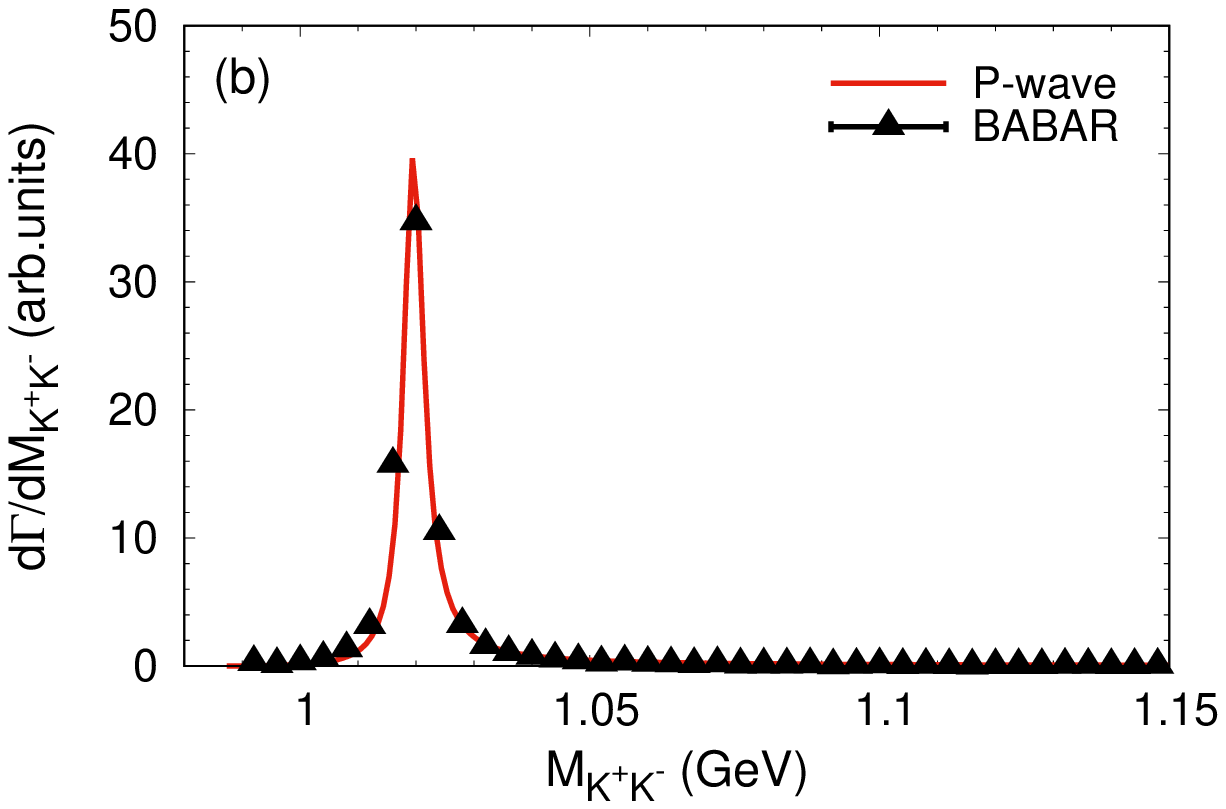}
  \caption{\small{The $K^+K^-$ mass distributions of the decay $D_s^+ \to K^+ K^- \pi^{+}$ in $P$-wave with the fitted parameters  of Table~\ref{table:p}. The BESIII data are taken from Fig.~9(b) of Ref.~\cite{Ablikim:2020xlq}, and  the {\it BABAR} data are taken from Fig.~5(a) of Ref.~\cite{delAmoSanchez:2010yp}.}}
  \label{fig:pp}
  \end{figure}

Next, we fit the parameter $\beta$ to the BESIII and {\it BABAR} measurements of the $K^+K^-$ mass distribution of the decay $D_s^+\to K^+K^-\pi^+$ in $P$-wave~\cite{Ablikim:2020xlq,delAmoSanchez:2010yp}. The fitted results are given in Table~\ref{table:p}. With the fitted parameters, in Fig.~\ref{fig:pp} we present the $K^+K^-$ mass distribution, together with the BESIII and {\it BABAR} measurements. One can find the clear peak of $\phi$ in the mass distributions, and our results are also in good agreement with the BESIII and {\it BABAR} measurements~\cite{Ablikim:2020xlq,delAmoSanchez:2010yp}.

\begin{figure}[htpb]
  \centering
  \includegraphics[scale=0.5]{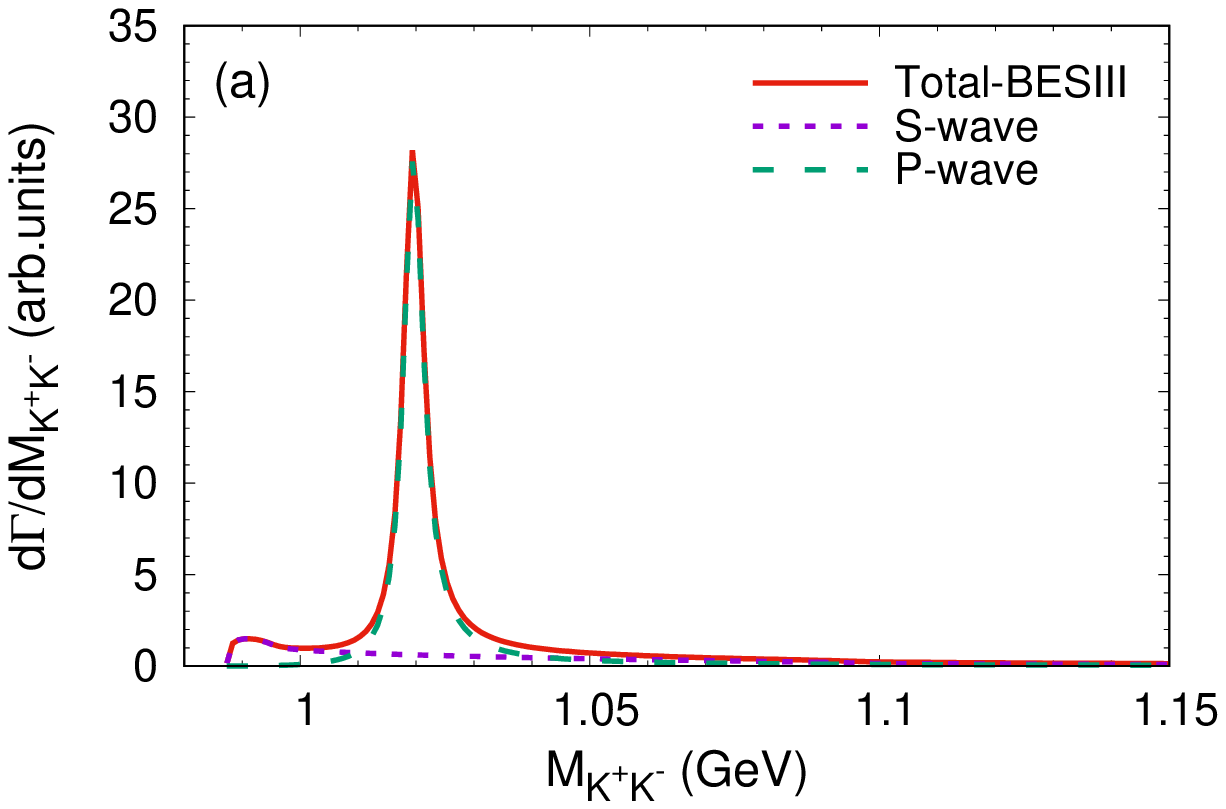}
  \includegraphics[scale=0.5]{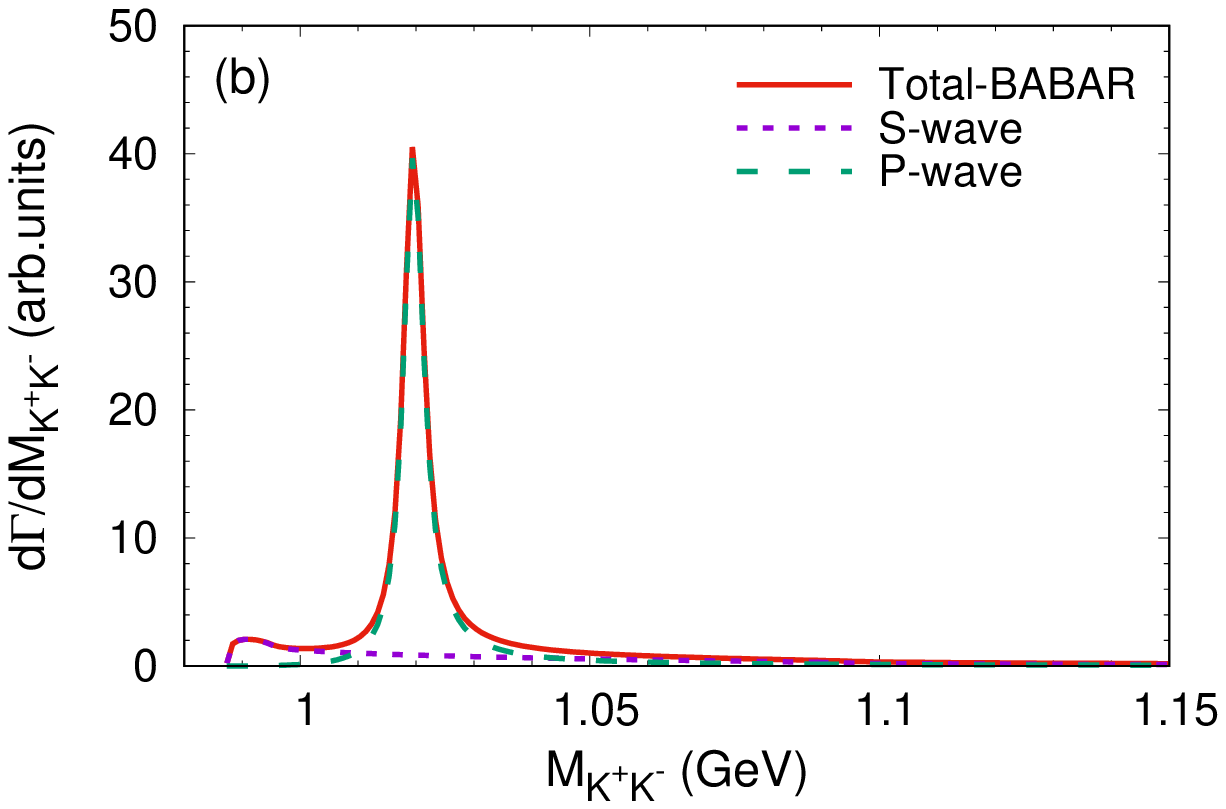}
  \caption{\small{The $K^+K^-$ mass distributions for the decay $D_s^+ \to K^+ K^- \pi^{+}$ by summing the contributions of $S$-wave and $P$-wave, (a) the results with the parameters derived from the BESIII measurements~\cite{Ablikim:2020xlq}, (b) the results with the parameters derived from the {\it BABAR} measurements~\cite{delAmoSanchez:2010yp}.}}
  \label{fig:total}
  \end{figure}

Finally, we also present the results by summing the contribution of $S$-wave and $P$-wave in Fig.~\ref{fig:total}, with the parameter derived from the BESIII and {\it BABAR} measurements. One can find the clear enhancement close to the 
$K^+K^-$ threshold, and our results are in agreement with the BESIII measurements (see Fig.~8(a) of Ref.~\cite{Ablikim:2020xlq}),  which implies that the $S$-wave pseudoscalar-pseudoscalar interaction plays an important role in the process $D_s^+\to K^+K^-\pi^+$, and the reasonable description of the $a_0(980)$ and $f_0(980)$ nature is crucial for us to interpret the experimental measurements. 

\section{Conclusions}
\label{sec:conc}

In this work, we have investigated the Cabibbo-favored process of $D_s^+ \to K^{+} K^{-} \pi^{+}$, by considering the $S$-wave pseudoscalar-pseudoscalar interation within the chiral unitary approach,  and  also the $P$-wave contribution from the intermediate vector meson $\phi$. For the decay $D_s^+ \to K^{+} K^{-} \pi^{+}$ in $S$-wave, in addition to the $W^+$ external emission mechanism considerd in Ref.~\cite{Dias:2016gou}, we also consider that $W^+$ internal emission mechanism. We  find that the isospin $I=1$ component of the $K^+K^-$ system is cancelled, and only the $I=0$ component contributes to the $S$-wave $D_s^+ \to K^{+} K^{-} \pi^{+}$.
By fitting to the $S$-wave and $P$-wave $K^+K^-$ mass distributions measured by the BESIII and {\it BABAR} Collaborations~\cite{Ablikim:2020xlq,delAmoSanchez:2010yp}, we can conclude,

1) The combined state $S(980)$ of $a_0(980)$ and $f_0(980)$ with the the lineshape of Eq.~(\ref{eq:bw}), used in the fit of the $S$-wave $K^+K^-$ mass distribution by the BESIII and {\it BABAR} Collaborations, is not suitable, since the  lineshape of Eq.~(\ref{eq:bw}) for these resonances  is not adequate to produce the amplitudes involved.  

2) We have shown that, even considering the $W^+$ internal emission mechanism,  the $f_0(980)$ resonance gives the dominant contribution to the enhancement structure near the $K^+K^-$ threshold, in agreement with the argument of Ref.~\cite{Dias:2016gou} and the analysis of the {\it BABAR}~\cite{delAmoSanchez:2010yp}. The $a_0(980)$ is expected to play a very small contribution in the process $D_s^+ \to K^{+} K^{-} \pi^{+}$,  although the decay $D_s^+\to a_0(980)^0\pi^+$ has been observed through $D_s^+\to \pi^+\pi^-\eta$.  It should be pointed out that $f_0(980)$ is also found to play the dominant role near the $K^+K^-$  threshold by investigating the $\bar{B}\to f_0(980)$ form factor in Ref.~\cite{Cheng:2019tgh}.

3) With the interpretation of  $f_0(980)$  as the resonance dynamically generated from the $S$-wave pseudoscalar-pseudoscalar interaction within the chiral unitary approach, we can provide an excellent description for the $S$-wave $K^+K^-$ mass distribution of the process $D_s^+ \to K^{+} K^{-} \pi^{+}$.

4) In addition to the two normalization factors $V_p$ and $\beta$, we only have one parameter $C$, corresponding to the relative weight of the $W^+$ external emission mechanism with respect to the $W^+$ internal emission mechanism. The value of $C$ in the region $2\sim 3$ does not affect the lineshape of the $S$-wave $K^+K^-$ mass distribution. Our results are calculated with a minimum input, which indicates that our interpretation of the BESIII and {\it BABAR} experimental data is reasonable.

\appendix 
\section{The $P$-wave amplitude of $D_s^+$ decay}
\label{app:pwave}
Up to an unknown constant $\tilde{\beta}$, the amplitude of the $D_s^+ \to \phi \pi^+\to K^+K^-\pi^+$ is given by~\cite{Toledo:2020zxj},
\begin{eqnarray}
\mathcal{M}^P &=&\tilde{\beta} \left(p_{D_s^+}+p_{\pi^+}\right)^\mu \left(-g_{\mu\nu}+\frac{q_\mu q_\nu}{m^2_\phi}\right) \nonumber \\
&& \left( p_{K^+}-p_{K^-}\right)^\nu \frac{1}{q^2-m^2_\phi+i m_\phi\Gamma_\phi} \nonumber \\
&=&\frac{-\tilde{\beta}}{q^2-m^2_\phi+i m_\phi\Gamma_\phi}\left[ \left(p_{D_s^+}+p_{\pi^+}\right)\cdot \left( p_{K^+}-p_{K^-}\right)  \right.\nonumber \\
&& \left. + \frac{\left(p_{D_s^+}+p_{\pi^+}\right)\cdot\left(p_{D_s^+}-p_{\pi^+}\right)  }{m^2_\phi} \right. \nonumber \\
&& \left. \times \left( p_{K^+}+p_{K^-}\right)\cdot \left( p_{K^+}-p_{K^-}\right) \right] \nonumber \\
&=&\frac{-\tilde{\beta}}{q^2-m^2_\phi+i m_\phi\Gamma_\phi}\left[ \left(p_{D_s^+}+p_{\pi^+}\right)\cdot \left( p_{K^+}-p_{K^-}\right)  \right.\nonumber \\
&& \left. + \frac{\left(m^2_{D_s^+}-m^2_{\pi^+}\right) \left( m^2_{K^+}-m^2_{K^-}\right)  }{m^2_\phi} \right] \nonumber \\
&=&\frac{-\tilde{\beta}}{q^2-m^2_\phi+i m_\phi\Gamma_\phi}\left[ \left(p_{D_s^+}+p_{\pi^+}\right)\cdot \left( p_{K^+}-p_{K^-}\right)  \right] \nonumber \\
\end{eqnarray}
where $p_{D_s^+}$, $p_{\pi^+}$,  $p_{K^+}$, and $p_{K^-}$ are the four-momenta of the mesons $D_s^+$, $\pi^+$, $K^+$, and $K^-$, respectively.  $q=p_{D_s^+}-p_{\pi^+} =p_{K^+}+p_{K^-}$. We will perform the calculation in the $K^+K^-$ rest frame, and take the direction of the $\pi^+$ as the $z$-axis. The four-momenta of the mesons are,
\begin{eqnarray}
p_{D_s^+} &=& ( E_{D^+_s}, 0, 0, \tilde{p}_\pi ), \nonumber \\ 
p_{\pi^+} &=& ( E_{\pi^+}, 0, 0, \tilde{p}_\pi ), \nonumber \\ 
p_{K^+} &=& ( E_{K^+}, \tilde{p}_{K^+}{\rm sin}\theta{\rm cos}\phi, \tilde{p}_{K^+}{\rm sin}\theta{\rm sin}\phi, \tilde{p}_{K^+}{\rm cos}\theta ), \nonumber \\ 
p_{K^-} &=& ( E_{K^-}, -\tilde{p}_{K^+}{\rm sin}\theta{\rm cos}\phi, -\tilde{p}_{K^+}{\rm sin}\theta{\rm sin}\phi, -\tilde{p}_{K^+}{\rm cos}\theta ), \nonumber \\ 
\end{eqnarray}
with 
\begin{eqnarray}
\tilde{p}_{\pi^+}&=&\frac{\lambda^{1/2}\left(M^2_{D^+_s}, M^2_{\pi^+}, M^2_{K^+K^-}\right)}{2M_{K^+K^-}},\\
\tilde{p}_{K^+}&=&\frac{\lambda^{1/2}\left(M^2_{K^+K^-}, M^2_{K^+},M^2_{K^-} \right)}{2M_{K^+K^-}},
\end{eqnarray}
and $E_{K^+}=E_{K^-}=\sqrt{m^2_{K^\pm}+ \tilde{p}^2_{K^+}}$ and $q^2=\left( p_{K^+}+p_{K^-} \right)^2 =M^2_{K^+K^-}$. Now we have,
\begin{eqnarray}
\mathcal{M}^P &=&\frac{-\tilde{\beta}\times 4 \tilde{p}_{\pi^+}\tilde{p}_{K^+} {\rm cos}\theta }{M^2_{K^+K^-}-m^2_\phi+i m_\phi\Gamma_\phi}.
\end{eqnarray}
The $P$-wave $K^+K^-$ mass distribution can be obtained by integrating the angles,
\begin{eqnarray}
\frac{{\rm d}\Gamma}{{\rm d}M_{K^+K^-}} &=& \int_0^\pi {\rm sin}\theta{\rm d}\theta \int^{2\pi}_0 {\rm d}\phi \frac{1}{(2\pi)^4}\frac{p_{\pi^+}\tilde{p}_{K^+}}{8m^2_{D_s^+}} \left|\mathcal{M}^P\right|^2 \nonumber \\
&=&\frac{1} {(2\pi)^4}\frac{p_{\pi^+}\tilde{p}_{K^+}}{8m^2_{D_s^+}} \left|\frac{-\tilde{\beta}\times 4 \tilde{p}_{\pi^+}\tilde{p}_{K^+} }{M^2_{K^+K^-}-m^2_\phi+i m_\phi\Gamma_\phi}\right|^2 \nonumber \\
&& \times \int_0^\pi {\rm cos}^2\theta{\rm sin}\theta {\rm d}\theta \int^{2\pi}_0 {\rm d}\phi \nonumber \\
&=&\frac{1} {(2\pi)^4}\frac{p_{\pi^+}\tilde{p}_{K^+}}{8m^2_{D_s^+}} |\frac{-\tilde{\beta}\times 4 \tilde{p}_{\pi^+}\tilde{p}_{K^+} }{M^2_{K^+K^-}-m^2_\phi+i m_\phi\Gamma_\phi}|^2 \times \frac{4\pi}{3}\nonumber \\
&=&\frac{1} {(2\pi)^3}\frac{p_{\pi^+}\tilde{p}_{K^+}}{4m^2_{D_s^+}} |\frac{\tilde{\beta}\times 4 \tilde{p}_{\pi^+}\tilde{p}_{K^+} }{M^2_{K^+K^-}-m^2_\phi+i m_\phi\Gamma_\phi}|^2 \times \frac{1}{3}.\nonumber  \\
\end{eqnarray}
Taking into account the $\beta$ is unknown constant, we can rewrie the $S$-wave $K^+K^-$ mass distribution as,
\begin{eqnarray}
\frac{{\rm d}\Gamma}{{\rm d}M_{K^+K^-}} &=& \frac{1} {(2\pi)^3}\frac{p_{\pi^+}\tilde{p}_{K^+}}{4m^2_{D_s^+}} |\mathcal{M}^P|^2,\\
\mathcal{M}^P &=& \frac{\beta\times \tilde{p}_{\pi^+}\tilde{p}_{K^+} }{M^2_{K^+K^-}-m^2_\phi+i m_\phi\Gamma_\phi},  
\end{eqnarray}
with $\beta=\frac{4}{\sqrt{3}} \tilde{\beta}$.

\section{the $K^+K^-$ mass distributions of $|S|^2$ and $|P|^2$}
\label{app:B}
According to the Partial Wave Analysis of the BESIII and {\it BABAR} Collaborations~\cite{Ablikim:2020xlq,delAmoSanchez:2010yp}, the amplitude $\mathcal{M}$ can be obtained,
\begin{eqnarray}
{\mathcal{M}}&=&\sum\limits_{l}\mathcal{M}_{l}Y_{l}(\cos\theta),\label{expansion}
\end{eqnarray}
where  $Y_{l}(\cos\theta)$ are the spherical harmonic functions,
\begin{eqnarray}
        Y_{0}^{0} &=&\sqrt{\frac{1}{4\pi}},\nonumber\\
        Y_{1}^{0} &=&\sqrt{\frac{3}{4\pi}}\cos\theta ,\nonumber\\
        Y_{2}^{0} &=&\sqrt{\frac{5}{16\pi}}\left(3{\cos\theta}^2-1\right).\label{yk0}
\end{eqnarray}
Since the $S$-wave and $P$-wave amplitudes are important in the low energy region of the  $K^+K^-$  mass distribution of the process $D_s^+\to K^+K^-\pi^+$, the amplitude  $\mathcal{M}$ can be rewritten as
\begin{eqnarray}
{\mathcal{M}}&=&\mathcal{M}^{S}Y_{0}(\cos\theta)+\mathcal{M}^{P}Y_{1}(\cos\theta)\label{expansion1},
\end{eqnarray}
where $\theta$ is the angle between the $K^{+}$ direction in the $K^{+}K^{-}$ rest frame and the prior direction of the $K^{+}K^{-}$ system in the $D_{s}^{+}$ rest frame. Then the amplitude squared is
\begin{eqnarray}
|{\mathcal{M}}|^2&=&(\mathcal{M}^{S})^2Y_{0}^2(\cos\theta) +(\mathcal{M}^{P})^2Y_{1}^2(\cos\theta) \nonumber \\
&&  +2{\rm Re}\left(\mathcal{M}^{S}\mathcal{M}^{P}\right)Y_{0}(\cos\theta)Y_{1}(\cos\theta)\nonumber \\
&=&a Y_{0}(\cos\theta)+bY_{1}(\cos\theta)+cY_{2}(\cos\theta)\label{expansion4},
\end{eqnarray}
where $a$, $b$, and $c$ are coefficients. So the angular distribution can be expanded using spherical harmonic functions,
\begin{eqnarray}
        \frac{dN}{d\cos\theta} = 2\pi\sum\limits_{k=0}^{L_{{\rm max}}}\left\langle Y_{k}^{0}\right\rangle Y_{k}^{0}(\cos\theta),\label{expansion5}
\end{eqnarray}
where $L_{{\rm max}} = 2 \ell_{{\rm max}}$, and $\ell_{{\rm max}}$ is the maximum orbital angular momentum quantum number, and the harmonic functions $Y_{k}^0(\cos\theta)$ satisfies
\begin{eqnarray}
        \int_{-1}^{1}Y_{k}^{0}(\cos\theta)Y_{j}^{0}(\cos\theta) d\cos\theta  = \frac{\delta_{kj}}{2\pi}.\label{expansion6}
\end{eqnarray}
Multiply Eq.~(\ref{expansion5}) by $Y_{k}^0(\cos\theta)$, and the integration can be obtained
\begin{align}
        &\int_{-1}^{1}\frac{dN}{d\cos\theta}Y_{k}^{0}(\cos\theta) d\cos\theta \nonumber\\
        =&\int_{-1}^{1}2\pi\sum\limits_{j=0}^{L_{{\rm max}}}\left\langle Y_{j}^{0}\right\rangle Y_{j}^0(\cos\theta)Y_{k}^0(\cos\theta)d\cos\theta \nonumber\\
        =& 2\pi\sum\limits_{j=0}^{L_{{\rm max}}}\frac{\delta_{kj}}{2\pi}\left\langle Y_{j}^{0}\right\rangle \nonumber\\
        =&\left\langle Y_{k}^{0}\right\rangle  \label{expansion7}.
\end{align}
In this work $D_s^+ \to K^{+} K^{-} \pi^{+}$, we only consider the contribution of $S$-wave and $P$-wave in $K^+K^-$, we can get
\begin{eqnarray}
        \frac{dN}{d\cos\theta} &=& 2\pi|S\,Y_{0}^{0}(\cos\theta)+P\,Y_{1}^{0}(\cos\theta)|^2 \nonumber\\
        &=& 2\pi\left[|S|^2{Y_{0}^{0}}^2(\cos\theta)+|P|^2{Y_{1}^{0}}^2(\cos\theta)\right.\nonumber \\
        &&\left. +2|S||P|\cos\phi_{SP}Y_{0}^{0}(\cos\theta)Y_{1}^{0}(\cos\theta)\right]  \nonumber\\  \label{expansion8}
\end{eqnarray}
and according to Eq.~(\ref{yk0}), Eq.~(\ref{expansion8}) can be rewritten as,
\begin{eqnarray}
        \frac{dN}{d\cos\theta}&=&2\pi\left[|S|^2\frac{1}{4\pi}+|P|^2\frac{1}{4\pi}+|P|^2\frac{1}{\sqrt{5\pi}}Y_{2}^{0}(\cos\theta)\right.\nonumber \\
        &&\left. +2|S||P|\cos\phi_{SP}\frac{1}{\sqrt{4\pi}}Y_{1}^{0}(\cos\theta)\right],   \label{expansion11}
\end{eqnarray}
where $\phi_{{SP}} = \phi_{{S}} - \phi_{{P}}$ is the phase difference between the $S$-wave and $P$-wave. Then we can get the values of $\langle Y_{0}^{0}\rangle$, $\langle Y_{1}^{0}\rangle$ and $\langle Y_{2}^{0}\rangle$,
\begin{eqnarray}
        \left\langle Y_{0}^{0} \right\rangle &=&\int_{-1}^{1} \frac{dN}{d\cos\theta}Y_{0}^{0}(\cos\theta) d\cos\theta \nonumber\\
        &=& \frac{1}{\sqrt{4\pi}}(|S|^2+|P|^2),\nonumber\\
        \left\langle Y_{1}^{0} \right\rangle &=& \frac{1}{\sqrt{4\pi}}(|S|^2+|P|^2)\times2|s||p|\cos\phi_{sp},\nonumber \\
        \left\langle Y_{2}^{0} \right\rangle &=& \frac{1}{\sqrt{5\pi}}|P|^2. \label{expansion9}
\end{eqnarray}

For the three-body decay, we have~\cite{Zyla:2020zbs}, 
\begin{eqnarray}
d\Gamma&=&\frac{1}{(2\pi)^5}\frac{1}{16M^2}\left| \mathcal{M} \right|^2 |p^*_1| |p_3| dm_{12} d\Omega^*_1 d\Omega_3,\nonumber \\
\end{eqnarray}
then,
\begin{eqnarray}
\frac{d\Gamma}{dm_{12}d{\rm cos}\theta_1^*}&=&\int d\phi_1^* \int d\Omega_3 \frac{1}{(2\pi)^5}\frac{|p^*_1| |p_3|}{16M^2}\left| \mathcal{M}^{S} +\mathcal{M}^{P} \right|^2 \nonumber \\
&=&\int d\phi_1^* \int d\Omega_3 \frac{1}{(2\pi)^5}\frac{|p^*_1| |p_3|}{16M^2}\left( \left| \mathcal{M}^{S} \right|^2 \right. \nonumber \\
&& \left. +\left|\mathcal{M}^{P} \right|^2 + 2 {\rm cos}\phi_{SP}\left|\mathcal{M}^{S} \right|\left|\mathcal{M}^{P} \right| \right). \label{eq:3bodydecay}
\end{eqnarray}
By comparing Eq.~(\ref{expansion11}) to Eq.~(\ref{eq:3bodydecay}), one can easily find,
\begin{eqnarray}
&& |S|^2  \propto  \frac{1}{(2\pi)^5}\frac{|p^*_1| |p_3|}{16M^2}\left| \mathcal{M}^{S} \right|^2, \\
&& |P|^2  \propto  \frac{1}{(2\pi)^5}\frac{|p^*_1| |p_3|}{16M^2}\left| \mathcal{M}^{P} \right|^2 ,
\end{eqnarray}
which implies that the $K^+K^-$ mass distributions of the $|S|^2$ and $|P|^2$ reported by BESIII and {\it BABAR} approximate to the $S$-wave and $P$-wave $K^+K^-$ mass distributions of the decay $D_s^+\to K^+K^-\pi^+$.

\begin{acknowledgments}
We warmly thank Eulogio Oset and Shuntaro Sakai for useful comments.
This work is partly supported by the National Natural Science Foundation of China under Grants No. 11505158.  It is also supported by the Key Research Projects of Henan Higher Education Institutions under No. 20A140027, the Project of Youth Backbone Teachers of Colleges and Universities of Henan Province (2020GGJS017), the
Natural Science Foundation of Henan (212300410123), the Fundamental Research Cultivation Fund for Young Teachers of Zhengzhou University (JC202041042), and the Academic Improvement Project of Zhengzhou University.
\end{acknowledgments}

\end{document}